\documentclass[12pt]{article}

\usepackage[english]{babel}
\usepackage{amsfonts}
\usepackage{amssymb}
\usepackage{graphicx}

\newcommand{\be}{\begin{equation}}
\newcommand{\ee}{\end{equation}}
\newcommand{\bea}{\begin{eqnarray}}
\newcommand{\eea}{\end{eqnarray}}

\renewcommand{\k}{\kappa}
\begin{document}

\begin{titlepage}

\begin{flushright}
{\tt
    hep-th/0601084}
 \end{flushright}

\bigskip

\begin{center}

{\Large \bf{Back-reaction effects in acoustic black holes}}

\bigskip
\bigskip\bigskip
S. Fagnocchi\footnote{fagnocchi@bo.infn.it}\footnote{Talk given at the
conference ``Constrained Dynamics and Quantum Gravity (QG05)", Cala Gonone (Italy), 
September 2005}

\end{center}

\bigskip%

\footnotesize \noindent {\it Centro Enrico Fermi, Compendio Viminale, 00184 Roma (Italy)\\ Dipartimento di Fisica dell' Universit\'a di Bologna and INFN sezione di Bologna, \\via Irnerio 46, 40126 Bologna (Italy) }

\bigskip

\bigskip
\bigskip \bigskip \bigskip 
\begin{center}
{\bf Abstract}
\end{center}
Acoustic black holes are very interesting non-gravitational objects which can be described by the geometrical formalism of General Relativity.  These models  can be useful to experimentally test  effects otherwise undetectable, as for example the Hawking radiation.
The back-reaction effects on the background quantities induced by the analogue Hawking radiation could be the key to indirectly observe it.\\
We briefly show how this analogy works and derive the backreaction equations for the linearized quantum fluctuations in the background of an acoustic black hole. A first order in $\hbar$ solution is given in the near  horizon region. It indicates that acoustic black holes, unlike Schwarzschild ones, get cooler as they radiate phonons. They show remarkable analogies with near-extremal Reissner-Nordstr\"om black holes.

\bigskip


\end{titlepage}

\newpage

\section{Introduction}
Everybody agrees on the fundamental importance  in Physics of experimental confermations of the theoretical predictions. Yet unfortunately some fields of Physics suffer an incredible lack of experimental data, which can be a serious obstacle for  their future developments. Among these is Quantum Field Theory in curved space (the study of quantum fields in presence of gravitational sources), whose most spectacular prediction -- the Hawking effect -- widely considered a milestome in theoretical Physics, so far has  no experimental evidences at all which could confirm its existence. Taking into account Quantum Mechanics, Hawking \cite{hawking} showed that black holes are unstable objects which thermally radiate particles at a temperature proportional to the horizon surface gravity $k$, i.e. $T_H=\frac{\hbar k}{2\pi \k_B c}$ \footnote{A hint of its fundamental origin can be given observing this formula: three universal constants ($h$ from Quantum Mechanics, $c$ from Special Relativity and  $G$, contained into $k$, from General Relativity), 
appear.} where $c$ is the speed of light and $\k_B$ the Boltzmann constant.  Because of its weakness, it is rather impossible to detect Hawking radiation from astrophysical observations: the emission temperature for a solar mass black hole would be $\sim 10^{-8}\ K$, far beyond the cosmic microwave background temperature of $2.7 \ K$. Moreover things become even worst if the mass of the hole is higher. On the other hand, also detection of Hawking radiation from primordial mini black holes seems problematic since inflation would have dilute them together with all the population of  primordial objects \cite{cosm}. \\
A key feature of Hawking radiation is its absolute independence from the dynamical details. In fact it is not directly linked to the Einstein equations, but only to the geometrical properties of space-time near the horizon. It is a pure kinematical effect,  so in principle it will occur whenever a space-time structure similar to that of a black hole appears.
\\So recently a gleam has been opened in indirect detecting Hawking radiation through the so called \emph{analogue models} of gravitational systems in Condensed Matter Physics \cite{unruh81, libro, review}. It seems possible to simulate in laboratory black hole-like objects thanks to the astonishing analogy which exists between sound propagation in an inhomogeneous fluid flow and light trapped in the gravitational field of a black hole. In the following we will focus the attention on a hydrodynamical flow in a nozzle, which is the simplest setting where the analogy can work. In this kind of systems, it should be possible to create in laboratory acoustic analogue of gravitational black  holes which shall thermally radiate phonons for the same reason that gravitational black holes emit Hawking radiation, providing a new context to detect, albeit indirectly, it.

\section{The analogy}
Let us consider a fluid
which is irrotational (i.e. the velocity $\vec v$ can be expressed as the
gradient of a potential $\psi$) and homentropic (i.e. the pressure
$P$ is a function of the density $\rho$ only). The equations of motions which describe its flow are the continuity and the Bernoulli equations 
\bea &\dot\rho + \vec \nabla\cdot
(\rho\vec v)=0 \, ,&\label{cont0}\\ 
& \dot\psi + \frac{1}{2}\vec v^2 +\mu(\rho)=0\, ,&\label{bern0}\eea
where $\mu(\rho)\equiv \frac{du}{d\rho}$, with $u(\rho)$ the internal energy.\\
Now, expanding the fields $\psi$, $\rho$, $\vec v$ around a background  solution of Eqs. (\ref{cont0}, \ref{bern0}) as $\psi=\psi_0 +\epsilon \psi_1 $, $\rho=\rho_0 +\epsilon \rho_1 $, $\vec v=\vec v_0 +\epsilon \vec v_1 $ up to $O(\epsilon^2)$, with $\epsilon$ a dimensionless expansion parameter, the first order in $\epsilon$ equations of motions read
\bea
&\dot\rho_1 + \vec \nabla\cdot
(\rho_1\vec v_0+\rho_0\vec v_1)=0 \, ,&\label{cont1}\\
&\dot\psi_1 + \vec v_0 \cdot \vec v_1 +
\frac{c^2}{\rho_0}\rho_1 =0\, ,&\label{bern1}\eea
where we have introduced the speed of sound $c$
as $c^2=\rho \frac{d\mu}{d\rho}|_{\rho_0}$.
Inserting $\rho_1$, extracted from Eq. (\ref{bern1}), into Eq. (\ref{cont1}) it yields
\be\label{eqpsi1}
-\partial_t \left[ \frac{\rho_0}{2c^2}
(\partial_t\psi_1 + \vec v_0\cdot \vec\nabla\psi_1)\right] +\vec \nabla
\cdot \left\{ \vec v_0 \left[-\frac{\rho_0}{c^2}(\partial_t\psi_1 + \vec
v_0\cdot  \vec\nabla \psi_1)\right] + \rho_0\vec \nabla\psi_1 \right\} =0\ . \ee
Since sound is the first order fluctuation of the quantities which
describe the mean flow, Eqs. (\ref{bern1}) and (\ref{eqpsi1})
describe the usual propagation of sound waves. \\ The crucial point is that Eq. (\ref{eqpsi1})
can be written as the four-dimensional d'Alambertian equation 
\be
\label{eqpsi1f1}
 \partial_{\mu}(\sqrt{-g}g^{\mu\nu}\partial_\nu \psi_1)=0 \, ,\ee 
provided one introduces an "acoustic  metric tensor"
\be \label{gmunu} g_{\mu\nu}\equiv\frac{\rho_0}{c}\left(
\begin{array}{cc}
-(c^2 -v_0^2)& -\vec v_0^T\\
-\vec v_0& \mathbf{1}
\end{array}\right)\ . \ee 
So the propagation of sound waves coincides with the propagation of a massless scalar field in a fictitious curved background given by the \emph{acoustic metric} $g_{\mu\nu}$. It is  worth  noting that this is not the real Minkowski space-time metric lying underneath the fluid flow, which is  completely non-relativistic ($|\vec v_0|\ll c_{light}$). $g_{\mu\nu}$ is the metric  the fluctuations "feel" and is due to the non-homogeneities of the fluid flow which, as it should be, modify and distorce the motion of sound waves \cite{volovik}.\\ Inspection of (\ref{gmunu}) reveals that the component $g_{00}$ vanishes for $|\vec  v|=c$. This fact suggests that not only a four-dimensional Lorentzian formulation of the sound propagation can be given, but also that this propagation, when the flow reaches the speed of sound, can be similar to that of light in  a black hole space-time. \\
\begin{figure}
\begin{center}
\includegraphics[angle=270,width=3.4in,clip]{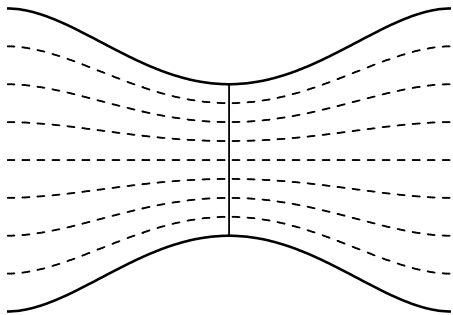}
\end{center}
\caption{\label{laval}Schematic representation of a de Laval nozzle. The fluid is supposed to flow from right to left, reaching the sound speed $c$ at the waist.}
\end{figure}
The simplest device to built this "acoustic black hole" is a de Laval nozzle , a converging-diverging nozzle (see Fig.\ref{laval}): let us suppose that the fluid flows  from right to left.  The flow can be adjusted to be subsonic in the right region, it accelerates reaching the sound velocity exactly at the waist, then  becomes supersonic into the left \cite{hydro}.  It is clear that for upstream sound waves propagating in the supersonic region of a de Laval nozzle is impossible to reach the subsonic region on the right of the nozzle. They are dragged by the flow and behave as light trapped in the gravitational field of a black hole. Therefore the supersonic side of the nozzle represents the sonic black hole and the waist the sonic horizon. In addition it is possible to show that the sound moves following the geodesics of the acoustic black hole metric (\ref{gmunu}). Now, pushing forward the analogy, after quantization of the field $\psi_1$ and using the same Hawking arguments, a flux of phonons (the sonic quanta of the field $\psi_1$) at the temperature $T_H$ is expected:
\be
T_H=\frac{\hbar k}{2\pi \k_B c}
\ee
where here $c$  is the speed of sound and $k$ is  the surface gravity calculated on the acoustic horizon defined as $ k=-\frac{1}{2}\left. \frac{d(c^2-v^2)}{dz}\right|_H$ \cite{unruh81, visser}.  This is the sonic analogue of Hawking radiation.   
In some cases this temperature is not so far from the actual experimental limits (see \cite{libro, bvl}).

\section{Back-reaction of phonons}
Unfortunately this flux of phonons, because of the extremely low associated emitted power ($\sim 10^{-31}\, W$ for water, $\sim 10^{-28}\, W$ for Bose-Einstein condensates), is still rather difficult to display; so it is of great interest to know how the phonon flux influences the background flow, i.e. to know its back-reaction effects, in order to have a chance to detect indirectly the Hawking radiation by the observation of its back-effects. But this is not a trivial extension of the analogy, since the back-reaction involves the dynamics of the system, while the analogy works only at the kinematical level. This issue can be addressed by considering the expansion of the equations of motion up to second order (i.e. taking $\psi=\psi_0 +\epsilon \psi_1 +\epsilon^2 \psi_2$, $\rho=\rho_0 +\epsilon \rho_1 +\epsilon^2 \rho_2$, $\vec v=\vec v_0 +\epsilon \vec v_1 +\epsilon^2 \vec v_2$).
The first order equations of motion (\ref{cont1}, \ref{bern1}) contain linear fluctuations (they describe the propagation of sound waves), while the back-reaction is given by the second order  equations, where the fluctuations appear quadratically:
\bea \dot\rho_2 + \vec\nabla \cdot(\rho_2\vec v_0+\rho_0\vec v_2) +
\vec\nabla \cdot(\rho_1\vec v_1) =0\ , \label{backeq1}\\
\dot\psi_2 + \vec v_0\cdot\vec v_2 + \mu(\rho_2) + \frac{1}{2}\vec
v_1^2 - \frac{c^2}{2\rho_0}\rho_1^2 =0\ ,\label{backeq2}\eea where we have assumed
for simplicity sake  $c=const$.\\ Making use of the standard definition of the stress-energy tensor\footnote{To be precise $T_{\mu\nu}$ is not the stress-energy tensor which gives the energy-momentum of the whole flux. It just refers to the energy-momentum content due to the fluctuations (i.e. the sound waves). For this reason it is usually called "pseudo" stress-energy tensor. For a detailed treatment see \cite{ volovik, stone}. } for a massless scalar field $ T_{\mu\nu}= \partial_\mu \psi_1\partial_\nu\psi_1
-\frac{1}{2}g_{\mu\nu}\partial^\alpha\psi_1\partial_\alpha\psi_1 $ and Eq. (\ref{bern1}), Eqs. (\ref{backeq1}, \ref{backeq2}) read (after quantum averaging over the fluctuations):
\bea & & \dot\rho_2 + \vec\nabla \cdot(\rho_2\vec v_0+\rho_0\vec v_2)
+\vec \nabla \cdot\left[ -\frac{\rho_B}{c^2}(\langle T_{ti}\rangle +v_B^j\langle T_{ij}\rangle )\right]=0 \ , \label{backeqstmunu1}\\
& & \dot\psi_2 + \vec v_0\cdot\vec v_2 + \mu(\rho_2) 
-\frac{1}{2}\frac{\rho_0}{c} \langle T\rangle  =0\ .
\label{backeqstmunu2}\eea where $T=g^{\alpha\beta}T_{\alpha\beta}$ is the trace of the
pseudo stress-energy  tensor and $\langle\; \rangle$ must be intended as quantum expectation values taken on the quantum state which properly describes the formation of a black hole, i.e. the Unruh state \cite{unruh76}. It is worth noting that,  although the dynamics for an hydrodynamical fluid flow deeply differs from the gravitational one, back-reaction effects are driven in both cases by the stress energy tensor, even if its components appear in different ways. Continuity equation, as one could have expected, acquires an extra term which takes into account the momentum-density of the phonons ($-\frac{\rho_B}{c^2}(T_{ti}+v_B^jT_{ij})=\rho_1 v_1^i=\sqrt{-g}T_i^0$). The Bernoulli equation instead, which provides the modification to the velocity, gets modified by the trace of the stress energy tensor. In the gravitational analogue equation is the flux which determines the mass correction. This is not surprising since one can not expect the very same back-reaction equations for the two settings, that, even if identical from the kinematical point of view, are completely different from the dynamical one \cite{ new, prl, prd}.

\section{An approximated solution}
A general solution for Eqs. (\ref{backeqstmunu1}, \ref{backeqstmunu2}) can not be given in a general four-dimensional setting, since the explicit form of the stress-energy tensor components is unknown. The problem can be partially addressed by introducing the approximation of quasi-one dimensional flow (i.e. the transverse -- let say $x,y$ -- components of the velocity are negligible with respect to the component along the  axes of the nozzle -- say $z$), which allows to perform a dimensional reduction. So  the two-dimensional back-reaction equations become
\bea \label{qev1}
&& A\dot{\rho_2}+\partial_z[A(\rho_0
v_2+\rho_2 v_0)]+\partial_z\left[-\frac{1}{c}\left(\langle
T_{tz}^{(2)}\rangle +v_0\langle T_{zz}^{(2)}\rangle\right)\right]=0\ , \\
\label{qev2}&& A\left(
\dot{\psi}_2+v_0\cdot v_2+\mu(\rho_2)\right)-\frac{1}{2}\langle
T^{(2)}\rangle=0. \eea
Here $A(z)$ is the area of the transverse section of the nozzle and $ T_{ab}^{(2)} \equiv A(\partial_a
\psi_1\partial_b \psi_1-\frac{1}{2}g^{(2)cd}\partial_c
\psi_1\partial_d \psi_1 g_{ab}^{(2)})$, with $g_{ab}^{(2)}$ the $(t,z)$-section of the four dimensional acoustic metric (\ref{gmunu}).\\ The quantum field $\psi_1$ now satisfies a two-dimensional equation for a massless scalar field coupled not only with the two-dimensional metric but also to the combination $A\rho/c$, which reminds the four-dimensional origin of the model: $\partial_a \left( \frac{A\rho}{c}\sqrt{-g^{(2)}}g^{ab}\partial_b \psi_1\right)=0$. This coupling causes backscattering of the modes. Now, neglecting this backscattering in accordance with the Polyakov approximation \cite{sandro}, the components of the stress-energy tensor are known for any general two-dimensional metric (see \cite{bd}).\\ We are interested in the first order in $\hbar$ correction to the mean flow, so the $\langle T_{ab}\rangle$ components must be evaluated on the background flow. Moreover we have chosen a shape for the transverse area of the nozzle as $A(z)=A_H(1+\kappa^2 z^2)$, which is  always  valid if one  restricts enough his analysis in a proper near horizon region. $A_H$ is the area at the waist, posed in $z=0$. For this particular choice for $A$, the horizon surface gravity becomes $k=c^2 \k$.\\
 The quantum corrections to the velocity and density are:
\bea
&&\epsilon^2v_2=\frac{\hbar}{A_H^2 \rho_0 e^{-1/2}c} (b_1 + c_1\k z)\k t\ ,\label{v2} \\
&&\epsilon^2 \rho_2=\frac{\hbar}{A_H^2 \rho_0 e^{-1/2}c}\frac{\rho_H e^{-1/2}}{c^2}(\alpha + \delta \k
z)t\ ,\label{r2}
\eea
where 
$b_1=9\gamma/2$,
$c_1=-304\gamma/15$, $\alpha=-a_2+c\k b_1$, $\delta=-b_2-a_2
+c\k c_1$, with $a_2=1189\gamma c\k/120$, $b_2=-161817\gamma
c\kappa/120$ and $\gamma\equiv \k c^2A_H/24\pi $.
These solutions  are only valid in the near horizon region ($\k z\ll 1$) and for small times ($c\k t\ll 1$). That means that they just give the hints of how the back-reaction starts modifying the background flow near the horizon. Inspection of Eqs. (\ref{v2}, \ref{r2}) reveals a depletion of both  velocity and density.\\
Then the position of the horizon, given by $|\vec v_{corrected}|=c$, moves: $z_H =- \frac{\epsilon b_1 t}{c}<0$,   shrinking the sonic black hole.\\ Also the emission temperature starts changing:
\be T=\left. \frac{\hbar}{2\pi\k_B}\frac{\partial
v}{\partial z}\right|_{z_H} = \frac{\hbar c}{2\pi\k_B}\k \left[ 1-
\frac{563}{720\pi}\epsilon \k^3 cA_H t\right]\ . \ee As time
goes on, the temperature of the emitted phonons decreases. Extrapolating this result supposing the system  passing  through stationary quasi-equilibrium states, one can try to get the late time evolution of $T_H$. What we have found is a lowering of $T_H\propto t^{-2}$: the sonic black hole temperature  vanishes in an infinite amount of time. This behavior completely differs from what one finds in Schwarzschild black holes, where the back-reaction increases the temperature and the black hole lifetime is of order $M_0^{-3}$($M_0$ is the initial mass of the hole). Acoustic black holes instead show a behavior similar to charged Reissner-Nordstr\" om black holes, whose temperature tends to zero as $\propto t^{-3}$ \cite{prl, prd}. 

\section{Comments}
The acoustic analogy in Condensed Matter Physics of gravitational objects is probably one of the most hopeful way to understand and verify the Quantum Field Theory (QFT) in black holes space-time (in particular Hawking radiation). Even if the first article by Unruh \cite{unruh81} is more than twenty years old, a lot of work has still to be done. In fact, no experimental proofs of Hawking radiation has been found yet, since the effects to point out are still quite tiny, even if not completely out of the present (or near future) experimental capability. \\
An easier way to  experimentally observe these kind of effects could be to appreciate the indirect presence of them via the modifications  they induce on the underlying system. So the study of back-reaction could be of primarily importance. In this work we have briefly presented the very first attempt in the analysis  of back-reaction in the simplest acoustic setting that provides a black hole analogue: a supersonic fluid flow in a de Laval nozzle. It displays that if the dynamics which drives the analogue system is correctly described by hydrodynamical equations, the back-reaction due to the Hawking radiation  is governed by the stress-energy tensor of the phonons.\footnote{This is not a general result: it holds only if the dynamics is  properly described by the hydrodynamical equations. For example for Bose-Einstein condensates it will not be true  \cite{schut}.}\\
A solution, under some quite reasonable approximations, tells us that the horizon should move shrinking in size the sonic black hole  and that the temperature emission  should drop  till to completely vanish in an infinite amount of time.\\ This is only a preliminary conclusion, that has to  be confirmed by numerical analysis and by further deeper studies, becoming the first step in the 
direction to fully make out the real prospects contained into the analogy.
\section*{Acknowledgements}
The author gratefully acknowledges the  Enrico Fermi Center for supporting her research.
 She also thanks R. Balbinot for useful comments and indications.


\end{document}